\documentstyle[twoside,fleqn,espcrc2]{article}

\newcommand{\Z}{{\sf Z \!\!\! Z}}

\newcommand{\AmS}{{\protect\the\textfont2
  A\kern-.1667em\lower.5ex\hbox{M}\kern-.125emS}}

\title{The confined-deconfined interface tension, wetting, and the spectrum
       of the transfer matrix}

\author{B. Grossmann$^{\small{\mbox{a}}}$,
M. L. Laursen\address{H\"ochstleistungsrechenzentrum, HLRZ c/o KFA J\"ulich,
P.O.Box 1913, 5170 J\"ulich, Germany}, T. Trappenberg\address{Institut f\"ur
Theoretische Physik E, RWTH Aachen,
Sommerfeldstrasse, 5100 Aachen, Germany} and U.-J. Wiese\address{Institut f\"ur
Theoretische Physik, Universit\"at
Bern, Sidlerstrasse 5, 3012 Bern, Switzerland}%
\thanks{Speaker at the conference, supported by Schweizer Nationalfond}}

\begin{document}

\begin{abstract}
The reduced tension $\sigma_{cd}$ of the interface between the confined and
the deconfined phase of $SU(3)$ pure gauge theory is determined from numerical
simulations of the first transfer matrix eigenvalues. At $T_c = 1/L_t$ we find
$\sigma_{cd} = 0.139(4) T_c^2$ for $L_t = 2$. The interfaces show universal
behavior because the deconfined-deconfined interfaces are completely wet by the
confined phase. The critical exponents of complete wetting follow from the
analytic interface solutions of a $\Z(3)$-symmetric $\Phi^4$ model
in three dimensions. We find numerical evidence that the confined-deconfined
interface is rough.
\end{abstract}

\maketitle

The tension of the interface between the low temperature hadron phase and the
high temperature quark-gluon plasma phase is an important parameter of the QCD
phase transition. In the early universe the interface tension determines the
nucleation rate of hadronic bubbles from the high temperature plasma. Thus, it
sets the scale for spatial inhomogeneities in the baryon density, which may
influence the primordial nucleosynthesis of light elements. To obtain an
estimate for the interface tension in QCD
we neglect the quarks and we concentrate
on an $SU(3)$ pure glue theory. Then the phase transition is first order, i.e.
the confined and the deconfined phase coexist at the critical temperature
$T_c$,
and the interface tension is nonzero. It is convenient to define the reduced
interface tension
\begin{equation}
\sigma_{cd} = \frac{F}{A T_c},
\end{equation}
where $F$ is the free energy of the interface and $A$ is its area.

In the pure glue theory the deconfinement phase transition is associated with
the spontaneous breakdown of the $\Z(3)$ center symmetry. Hence, as opposed to
full QCD, there are three distinct high temperature phases, which coexist at
temperatures $T > T_c$. Two different deconfined phases are separated by
deconfined-deconfined interfaces with a reduced interface tension
$\sigma_{dd}(T)$. At $T_c$ all four phases (three deconfined and one confined)
coexist with each other. Thermodynamical stability requires $\sigma_{dd}(T_c)
\leq 2 \sigma_{cd}$. Frei and Patk\'{o}s \cite{Fre89} have suggested that
\begin{equation}
\sigma_{dd}(T_c) = 2 \sigma_{cd}.
\end{equation}
Then a deconfined-deconfined interface consists of two confined-deconfined
interfaces with a macroscopic layer of confined phase in between. One says that
the deconfined-deconfined interfaces are completely wet by the confined phase.
Complete wetting is a critical phenomenon of interfaces (hence $T_c$ is really
a critical temperature although the bulk transition is first order). As one
approaches the phase transition from above the thickness of the confined
wetting layer
\begin{equation}
z_0 \propto (T - T_c)^{- \psi}
\end{equation}
diverges with a critical exponent $\psi$, while the order parameter at the
interface
\begin{equation}
\Phi_1(0) \propto (T - T_c)^\beta
\end{equation}
vanishes with another critical exponent $\beta$.

In our case the order parameter is the Polyakov loop
\begin{equation}
\Phi(\vec{x}) = \mbox{Tr} [{\cal P} \exp \int_0^{L_t} dt A_4(\vec{x},t)],
\end{equation}
where $A_\mu(\vec{x},t)$ is the anti-hermitean $SU(3)$ gauge potential.
The Polyakov loop $\Phi(\vec{x}) =
\Phi_1(\vec{x}) + i \Phi_2(\vec{x})$
is a complex scalar field in three dimensions. To compute the values of
the critical exponents one may construct an effective $\Z(3)$ symmetric
$\Phi^4$
theory with an action \cite{Tra92}
\begin{eqnarray}
&&S[\Phi] = \int d^3x [\frac{1}{2} \partial_i \Phi^* \partial_i \Phi +
V(\Phi)],
\nonumber \\
&&V(\Phi) = a |\Phi|^2 + b \Phi_1(\Phi_1^2 - 3 \Phi_2^2) + c |\Phi|^4.
\end{eqnarray}
The bare interfaces are solutions of the corresponding classical
equations of motion. Fig.1 shows the shapes $\Phi_1(z)$ and $\Phi_2(z)$ of a
planar deconfined-deconfined interface perpendicular to the $z$-direction deep
in the deconfined phase (a), in a region where wetting sets in (b), and very
close to the phase transition (c) where interface splitting and hence complete
wetting is clearly visible.
\begin{figure}[htb]
\vspace{75mm}
\caption{The deconfined-deconfined interface. The solid line
is $\Phi_1$ and the dashed line is $\Phi_2$.}
\end{figure}
Close to the phase transition an analytic interface solution yields
$z_0 \propto \log(T - T_c)$ and $\Phi_1(0) \propto \sqrt{T - T_c}$.
In particular,
the thickness $z_0$ of the confined complete wetting layer diverges
only logarithmically. The corresponding critical exponents are \cite{Tra92}
\begin{equation}
\psi = 0, \,\,\, \beta = \frac{1}{2}.
\end{equation}

Complete wetting is a peculiarity of the pure glue theory. In full QCD it does
not occur because the quarks break the $\Z(3)$ symmetry explicitly, thus
eliminating two of the three deconfined phases. In numerical simulations of the
pure glue theory wetting complicates the situation and makes the determination
of the reduced interface tension $\sigma_{cd}$ more difficult. To avoid the
complications it is advantageous to impose $C$-periodic boundary conditions
\cite{Kro91}. Just like quarks they break the center symmetry and they reduce
the number of deconfined phases from three to one \cite{Wie92}. With
$C$-periodic boundary conditions the Polyakov loop is replaced by its charge
conjugate when it is shifted a distance $L_i$ in the direction of the spatial
unit vector $\vec{e}_i$
\begin{equation}
\Phi(\vec{x} + L_i \vec{e}_i) = \,^C\Phi(\vec{x}) = \Phi^*(\vec{x}).
\end{equation}
Hence, the two deconfined phases with non-real expectation values of the
Polyakov loop are not consistent with $C$-periodic boundary conditions.
Of course,
one must use periodic boundary conditions in the euclidean time direction.

To extract the value of $\sigma_{cd}$ we investigate the system in a
cylindrical
spatial volume with two short $x$- and $y$-directions of lengths $L_x$ and
$L_y$ and a much longer $z$-direction. Close to $T_c$ typical
configurations then consist of several blocks of confined and deconfined
phases,
aligned along the $z$-direction, and separated by confined-deconfined
interfaces of area $A = L_x L_y$ spanned in the short $x$- and $y$-directions.
The interface shape $\Phi_1(z)$ (averaged over the short $x$- and
$y$-directions) is shown in fig.2 for a typical configuration on an
$8\times8\times128\times2$ lattice with $C$-periodic boundary conditions.
\begin{figure}[htb]
\vspace{35mm}
\caption{A typical interface shape.}
\end{figure}
The interfaces influence the spectrum of the transfer matrix in the
$z$-direction. In particular, the first two transfer matrix eigenvalues
$t_{0,1}(z) = \exp(- E_{0,1} z)$ can be obtained from a $2\times2$ matrix
of transition amplitudes
\begin{eqnarray}
t(z) = \left(\begin{array}{cc} t_{dd}(z) & t_{cd}(z) \\
t_{cd}(z) & t_{cc}(z) \end{array} \right).
\end{eqnarray}
To lowest order of the dilute interface approximation
the amplitude for transitions between
the confined and the deconfined phase is given by the one-interface
contribution
\begin{eqnarray}
&&t_{cd}(z) = \int_0^z dz_0 \nonumber \\
&&\exp(- x z_0) \delta \exp(- \sigma_{cd} A)
\exp(x (z - z_0)).
\end{eqnarray}
One integrates over the position $z_0$ of the interface. The Boltzmann factors
$\exp(- x z_0)$ and $\exp(x (z - z_0))$ represent a block of confined phase
of thickness $z_0$ and a block of deconfined phase of thickness $z - z_0$,
where $x = \frac{1}{2}(f_c - f_d) A/T$ depends on the free energy densities
$f_c$ and $f_d$ of the confined and the deconfined phase. The interface
contribution $\delta \exp(- \sigma_{cd} A)$ is suppressed by the interface
free energy $F/T = \sigma_{cd} A$, and it also contains a factor $\delta$ which
describes capillary wave fluctuations of the interface. A rigid interface
without capillary wave fluctuations has $\delta = 1$. Using similar expressions
for the amplitudes $t_{dd}(z)$ and $t_{cc}(z)$ and working to all orders of
the dilute interface approximation one finds an energy difference
$E_1 - E_0 = 2 \sqrt{x^2 + \delta^2 \exp(- 2 \sigma_{cd} A)}$,
which is minimal
for $x = 0$ \cite{Gro92}. Putting $E_0 = 0$ one obtains at the minimum
\begin{equation}
E_1 = 2 \delta
\exp(- \sigma_{cd} A).
\label{Cperiodic}
\end{equation}
In the numerical simulations the energy difference is
determined from the exponential decay of a correlation function of Polyakov
loops separated in the $z$-direction. We have used lattices $4\times4\times64$,
$4\times6\times64$, $6\times6\times64$, $6\times8\times96$ and
$8\times8\times 128$ always with $L_t = 2$ lattice points in the euclidean
time direction, and with several values of the Wilson coupling very close to
$\beta_c = 5.0933(7)$. Fig.3a shows the energy $E_1$ for simulations on a
$4\times6\times64\times2$ lattice with $C$-periodic boundary conditions. As
expected the minimum of $E_1$ occurs close to $\beta_c$, where $f_c = f_d$
and hence $x = 0$.
\begin{figure}[htb]
\vspace{70mm}
\caption{The spectra of the lowest energies.}
\end{figure}
Fig.4 shows a fit of $\log E_1$ as a function of the
interface area $A = L_x L_y$.
\begin{figure}[htb]
\vspace{55mm}
\caption{The dependence of the energies on the interface area $A$.}
\end{figure}
According to eq.(\ref{Cperiodic}) the slope of the curve
determines the value of the reduced interface tension $\sigma_{cd} = 0.035(1) =
0.140(4) T_c^2$. The fluctuation factor $\delta = 0.196(5) \neq 1$
indicates that the
confined-deconfined interface is rough as it should be in the continuum limit.
Note that rigid interfaces are lattice artifacts.

When periodic boundary conditions are used the situation becomes more
complicated, because then three deconfined phases are present and wetting may
arise. In particular, there are four energies $E_0$, $E_1$, $E_2$ and $E_3$
which become degenerate at $T_c$ in the infinite volume limit. In case of
complete wetting (putting again $E_0 = 0$) we find
$E_{1,2} = \sqrt{x^2 + 3 \delta^2 \exp(- 2 \sigma_{cd} A)} - x$ and
$E_3 = 2 \sqrt{x^2 + 3 \delta^2 \exp(- 2 \sigma_{cd} A)}$ \cite{Gro92}.
The energies $E_1$ and $E_2$ are
degenerate because of $\Z(3)$ symmetry and charge conjugation invariance. The
energy $E_3$ has its minimum at $x = 0$ where
\begin{equation}
E_3 = 2 \sqrt{3} \delta \exp(- \sigma_{cd} A) = 2 E_{1,2}.
\label{periodic}
\end{equation}
On the other hand, if wetting is incomplete (i.e. if $\sigma_{dd}(T_c) <
2 \sigma_{cd}$) the minimum of $E_3$ occurs at $x = - \gamma
\exp(- \sigma_{dd} A)
< 0$, while the point where $E_3 = 2 E_{1,2}$ is shifted to $x = 2 \gamma
\exp(- \sigma_{dd} A) > 0$ \cite{Gro92}.
The transfer matrix spectrum is then qualitatively
different. Fig.3b shows the spectrum on a $6\times6\times64\times2$ lattice
with periodic boundary conditions. The minimum of $E_3$ and the point where
$E_3 = 2 E_{1,2}$ are only slightly
shifted away from each other. In case of incomplete wetting one would expect
a shift, but it would go exactly in the opposite direction. Therefore,
our data are consistent with complete and not with incomplete
wetting. In fact, using eq.(\ref{periodic}) for an independent fit which is
also shown in fig.4, we obtain $\sigma_{cd} = 0.034(2) = 0.136(8) T_c^2$ and
$\delta = 0.202(6)$ in agreement with the $C$-periodic results.

To summarize we have found evidence that the deconfined-deconfined interfaces
are completely wet by the confined phase.
Combining the $C$-periodic with the periodic results we obtain
\begin{equation}
\sigma_{cd} = 0.139(4) T_c^2
\end{equation}
for the reduced confined-deconfined interface tension. Our
result is consistent with the one of the Boston group $\sigma_{cd} = 0.12(2)
T_c^2$ \cite{Hua90} and larger
than the one of the Helsinki group $\sigma_{cd} = 0.08(2) T_c^2$ \cite{Kaj90}.
As opposed to these methods we use the necessarily present finite size effects
to extract the value of $\sigma_{cd}$, instead of trying to avoid finite size
effects by going
to larger lattices. In particular, a reliable extrapolation to the infinite
volume limit is then possible, because we find agreement with our analytic
finite size formulae. The fluctuation factor $\delta = 0.199(4)$ indicates
that the confined-deconfined interface is rough, as it should be in the
continuum limit. Of course,
our results are obtained with $L_t = 2$ and it would be interesting to get
closer to the continuum limit, e.g. to $L_t = 4$. This is clearly feasible
using our finite size technology, but it requires larger lattices like e.g.
$16\times16\times256\times4$ because the phase transition is then more weakly
first order. Also high statistics is needed to be able to determine reliably
the spectrum of the transfer matrix. The existing data already indicate that
the interface tension is small at least in the pure glue theory. If this
remains
true in full QCD, only small spatial inhomogeneities are produced at the QCD
phase transition and primordial nucleosynthesis is likely to take place under
spatially homogeneous conditions.

\end{document}